\documentclass[prl,twocolumn,superscriptaddress,letterpaper,showpacs,floatfix,ltxgridinfo]{revtex4}
\usepackage{amsmath}
\usepackage{graphicx}
\usepackage{bibunits}

\newcommand{\sfrac}[2]{{^{#1}}\!/_{#2}}

\newcommand{\w}{\omega}
\renewcommand{\k}{\kappa}
\newcommand{\wm}{\w_m}

\newcommand{\ws}{\w_s}
\newcommand{\wcav}{\w_\mathrm{cav}}
\newcommand{\zho}{z_\mathrm{HO}}
\newcommand{\Z}{\mathcal{Z}}
\newcommand{\Zho}{\Z_\mathrm{HO}}
\newcommand{\fmNum}{110~\mathrm{kHz}}
\newcommand{\dca}{\Delta_{ca}}

\newcommand{\dpc}{\Delta_{pc}}

\newcommand{\nubar}{\bar{\nu}}
\newcommand{\nbar}{\bar{n}}
\newcommand{\Com}{C_\mathrm{om}}

\newcommand{\Gm}{\Gamma_m}

\newcommand{\gom}{g_\mathrm{om}}

\newcommand{\wbw}{\w_\mathrm{BW}}

\newcommand{\dop}[1]{#1^\dagger}

\newcommand{\tb}{\tilde b}

\newcommand{\placefigs}{
	\begin{figure*}[t]
	\centering
  \includegraphics[scale=0.9]{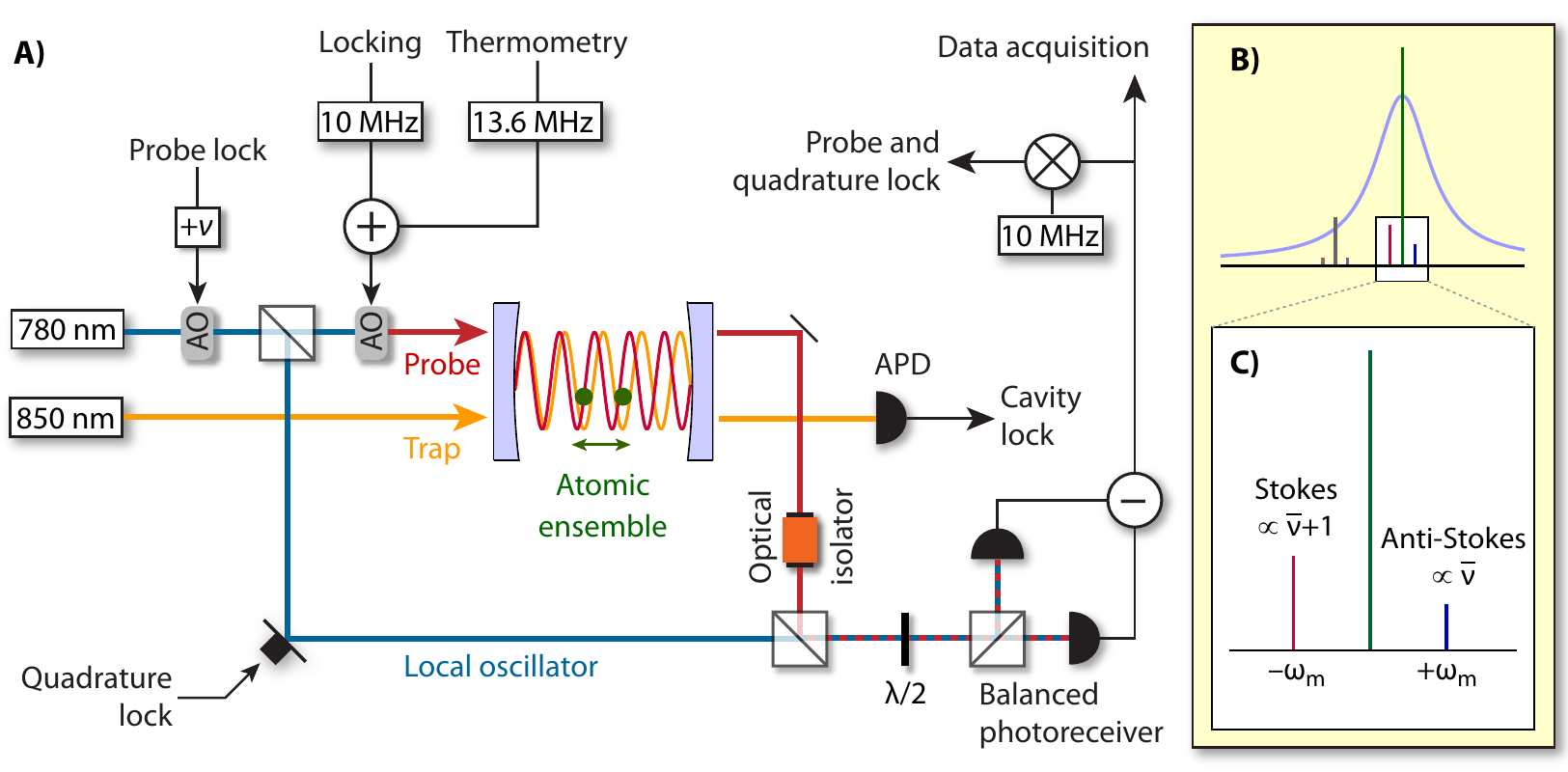}
	\caption{Schematic of the experiment.  (A) An ensemble of 4,000 ultracold $^{87}$Rb atoms is trapped in an optical standing wave within a high-finesse optical cavity.  The ensemble's center-of-mass motion coherently scatters light from cavity probe light, detected using a balanced heterodyne receiver.  (B) Probe spectrum in the cavity.  A strong resonant tone (green) applies backaction to and acquires sidebands from the collective atomic motion.  A weak detuned tone (gray) is used for locking the probe frequency with respect to the cavity, and does not significantly affect the oscillator, neither via incoherent backaction nor via dynamical cooling.  (C) A harmonic oscillator in its ground state can only extract energy from the optical field, leading to an asymmetry in the resonant probe's Raman sidebands, indicative of the mean phonon occupation number $\nubar$.
	}
	\label{fig:apparatus}
\end{figure*}

\begin{figure}[t!]
	\centering
		\includegraphics[width=\linewidth]{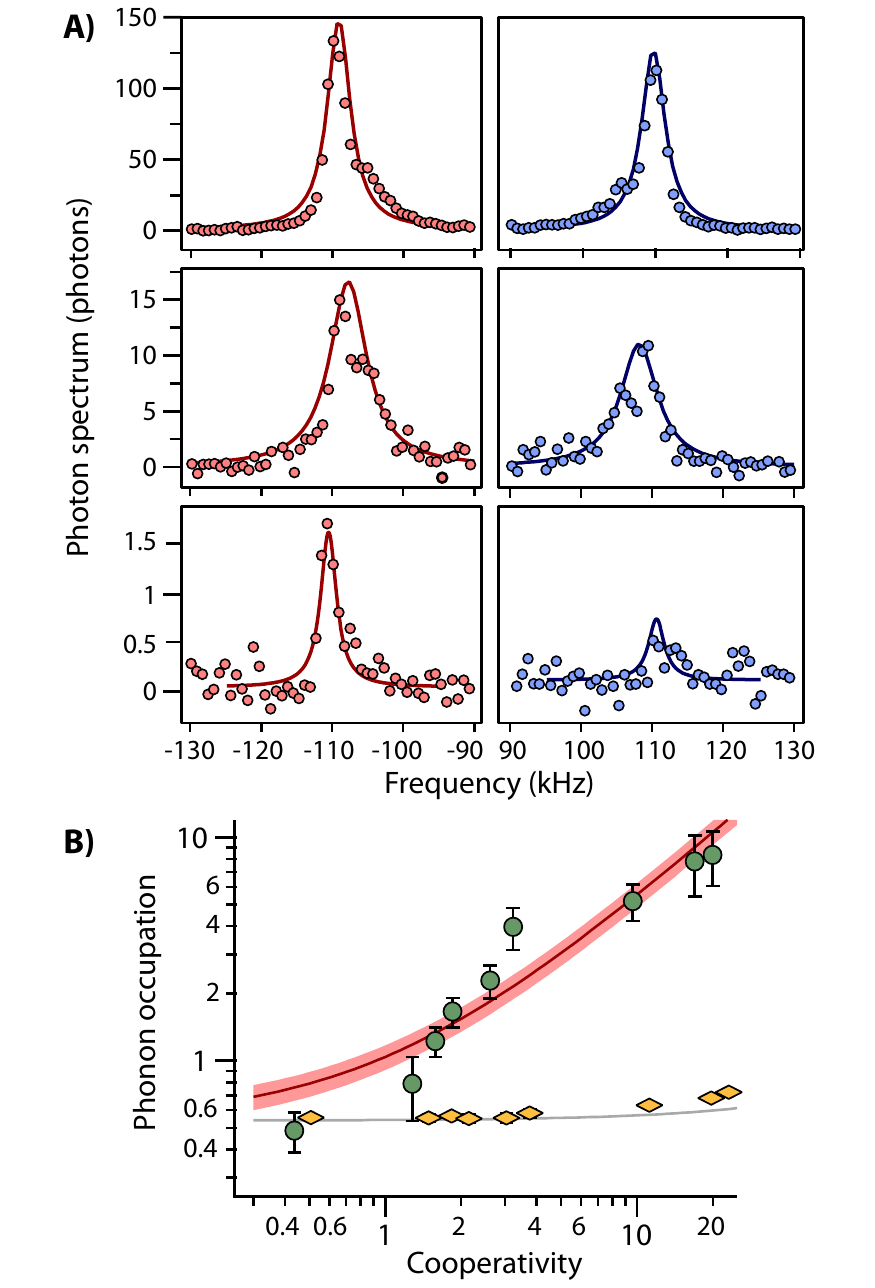}
	\caption{Asymmetric optical scattering from quantum collective motion.  (A) Photon spectrum of light exiting the cavity, scattered by collective atomic motion.  Shown are the Stokes sidebands (left panels, red) and anti-Stokes sidebands (right panels, blue) at various cooperativities (from top to bottom, $\Com=9.6,\,1.9,\,0.4$), together with the prediction of Eqn.~(\ref{eq:psd}), plotted \textit{vs.} cyclic frequency $f=\w/2\pi$.  The mechanical linewidths and resonance frequencies are fit to the data, but the peak heights are not free parameters. (B) Measured phonon occupation \textit{vs.} cooperativity.  The collective mode occupation (green circles) increases by $\Com/2$, according to the zero-free-parameter measurement backaction theory (red line, shaded region indicates 68\% systematic confidence interval).  In contrast, the r.m.s.\ single-atom axial occupation, measured using time-of-flight thermometry of the gas (yellow diamonds, measured at 1 ms of probing), remains near its initial value during the measurement (gray line indicates theoretical prediction).
	}
	\label{fig:thermometry}
\end{figure}

\begin{figure}[t]
	\centering
		\includegraphics[width=\linewidth]{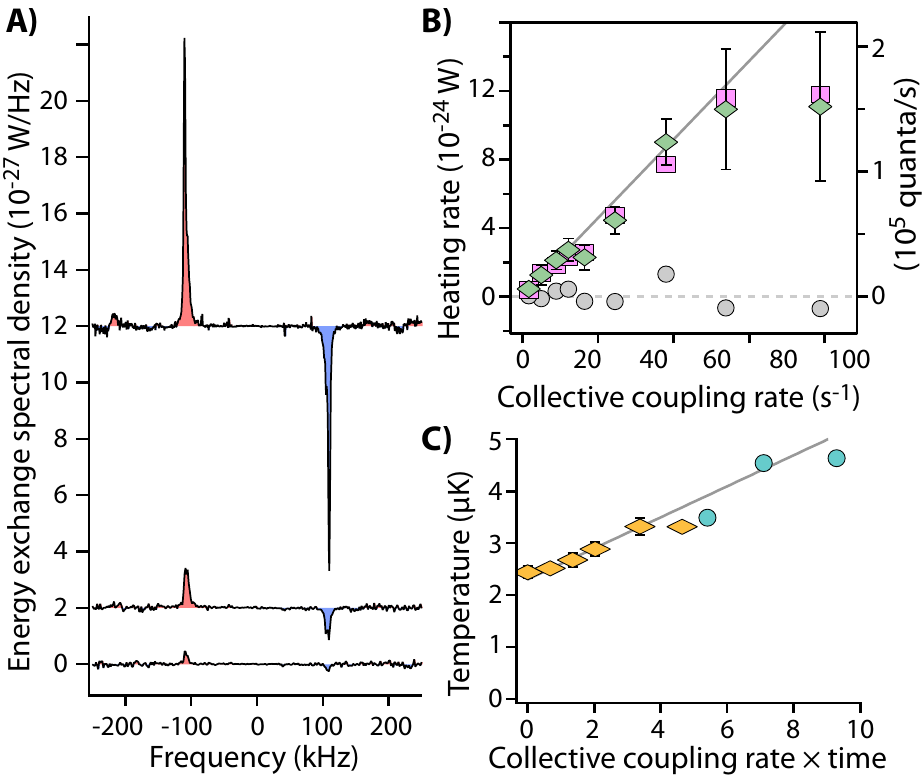}
	\caption{Energy exchange between light and motion.  (A)  Energy-exchange spectral densities for various probe cooperativities (from top to bottom, $\Com = 9.6,\, 3.2,\, 1.6$).  Traces are vertically offset for clarity (by $12,\,2,\,0\mathop\times 10^{-27}\,\mathrm{W/Hz}$).  (B)  Heating rate due to the resonant probe (magenta squares), the weak locking tone (gray circles), and the summed heating from both tones (green diamonds), together with zero-free-parameter measurement-backaction theory (gray solid line), plotted as a function of the collective coupling rate $4\nbar \gom^2/\k = \Gm \Com$.  (C) Temperature of the atomic gas \textit{vs.} the product of the collective coupling rate and the total interrogation time (yellow diamonds at 5~ms, cyan circles at 9~ms).  Also shown is the theoretical increase in the bath temperature (gray line).
	}
	\label{fig:exchange}
\end{figure}
}

\begin{document}

\title{Optically detecting the quantization of collective atomic motion}
\author{Nathan Brahms}
\email{nbrahms@berkeley.edu}
\author{Thierry Botter}
\author{Sydney Schreppler}
\author{Daniel W.C.\ Brooks}
\affiliation{Department of Physics, University of California, Berkeley, CA 94720, USA}
\author{Dan M.\ Stamper-Kurn}
\affiliation{Department of Physics, University of California, Berkeley, CA 94720, USA}
\affiliation{Materials Sciences Division, Lawrence Berkeley National Laboratory, Berkeley, CA  94720, USA}
\date{\today}

\begin{abstract}
We directly measure the quantized collective motion of a gas of thousands of ultracold atoms, coupled to light in a high-finesse optical cavity.  We detect strong asymmetries, as high as 3:1, in the intensity of light scattered into low- and high-energy motional sidebands.   Owing to high cavity-atom cooperativity, the optical output of the cavity contains a spectroscopic record of the energy exchanged between light and motion, directly quantifying the heat deposited by a quantum position measurement's backaction.  Such backaction selectively causes the phonon occupation of the observed collective modes to increase with the measurement rate.  These results, in addition to providing a method for calibrating the motion of low-occupation mechanical systems, offer new possibilities for investigating collective modes of degenerate gases and for diagnosing optomechanical measurement backaction.
\end{abstract}

\maketitle
\thispagestyle{plain}

\begin{bibunit}[apsrev4-1]
Quantized motion leads to a large asymmetry in the spectrum of light scattered by a ground-state oscillator.  Such asymmetry is most commonly observed from microscopic oscillators, such as electrons bound within atoms and molecules, single neutral atoms \cite{jessen:prl:92}, or few-ion ensembles \cite{diedrich:prl:89,islam:nc:11,monz:prl:11}.  In contrast, quantum aspects of the motion of massive, many-atom oscillators are typically obscured by thermal noise and high phonon occupation.  These objects, therefore, usually modulate the spectrum of light in a classical, symmetric manner.

Controlling and measuring the motion of macroscopic objects at levels sensitive to quantum effects will be critical for operating gravitational-wave detectors \cite{goda:nphys:08}, verifying the correspondence principle at macroscopic scales \cite{marshall:prl:03,romero-isart:prl:11}, and realizing protocols that mechanically store and exchange quantum information \cite{stannigel:prl:10,lin:nphot:10}.  Such goals are being pursued actively using cavity optomechanical systems, wherein the motion of an object with mass ranging from attograms to kilograms is observed via its coupling to an electromagnetic cavity \cite{kippenberg:s:08}.
Implementations of cavity optomechanics using the collective motion of atomic gases \cite{gupta:prl:07,brennecke:s:08} have demonstrated sensitivity to quantum optical force fluctuations, leading to the observation of optical squeezing from ponderomotive interactions \cite{brooks:11} and backaction from a quantum-limited position measurement \cite{murch:nphys:08}. 
\placefigs

In this work, we use a high-finesse optical cavity to detect the coherent, asymmetric scattering of light by collective modes of motion of a trapped atomic gas, occupied with as few as 0.5 phonons.  Observations have previously been made of optical emission asymmetries from individual atoms \cite{jessen:prl:92} and ensembles of up to 14 ions \cite{diedrich:prl:89,islam:nc:11,monz:prl:11}, and of asymmetric absorption by a nanomechanical solid-state resonator \cite{safavi-naeini:11}.  In this work, we measure the coherent scattering of light from the collective motion of thousands of ultracold atoms.  The scattering asymmetry acts as a self-calibrating thermometer for the atoms' collective phonon occupation.  Moreover, owing to high cavity-atom cooperativity and thermal isolation in our system, the cavity mode acts as the dominant channel for energy flux to our mechanical system.  The spectrum of light emitted from the cavity therefore serves as a record of the energy exchanged between motion and the light field.  We demonstrate that this energy transfer represents the necessary minimum diffusive heating of a continuous backaction-limited quantum position measurement.

Our experiment (Fig.~\ref{fig:apparatus}) begins with an ensemble of 4,000 ultracold $^{87}$Rb atoms.  The atoms are trapped in a few adjacent minima of a one-dimensional optical standing-wave potential, formed by 850-nm-light light resonating within a high-finesse Fabry-P\'erot optical cavity, and detuned far from atomic resonance.  The curvature at each potential minimum corresponds to an oscillation frequency, along the cavity axis, of $\wm = 2\pi\mathop\times\fmNum$.

We create an optomechanical coupling, which is linearly sensitive to atomic position, by trapping the atoms at locations with strong intensity gradients of 780-nm-wavelength probe light, which also resonates within the cavity \cite{purdy:prl:10}.  The probe light is detuned from the atomic D2 transition by an amount $\dca$, ranging from $-2\pi \mathop\times 12\textrm{ to }70~\mathrm{GHz}$.  At such detunings, the atomic gas acts as a position-dependent refractive medium, leading to an interaction Hamiltonian $\hat H_\mathrm{int} = \sum_i g_i \hat n \hat z_i/\zho$, where $\hat n$ is the probe's photon number operator and $\hat z_i$ is the position operator of atom $i$ \cite{murch:nphys:08}.  Here $g_i$ represents the change in the cavity's resonance frequency as atom $i$ is displaced by one single-atom harmonic oscillator length $\zho$, equivalent to $\sqrt{\hbar / 2 m \wm}$ for rubidium mass $m$.

The above may be rewritten \cite{gupta:prl:07,brennecke:s:08} as a collective interaction described by $\hat H_\mathrm{int} = \gom \hat n \hat \Z / \Zho$, where $\hat \Z \equiv \sum_i g_i \hat z_i/\sum_i g_i$ is a collective position operator, $\Zho = \zho/\sqrt{N_a}$ is the collective oscillator length for $N_a$ atoms, and $\gom = \sum_i g_i/\sqrt{N_a}$ is the collective optomechanical coupling rate.  In our system, the $g_i$ are all approximately equal, and $\hat \Z$ is nearly equivalent to the center-of-mass position $\hat \Z_\mathrm{com} = \sum_i \hat z_i/N_a$.  In this case, $\gom = \sqrt{N_a} g_0^2 k \zho / \dca$, with single-atom cavity-QED coupling rate $g_0$ ($2\pi\mathord\times 12.4~\mathrm{MHz}$, averaged over the intracavity atomic distribution) and probe wavenumber $k$.

Unlike solid-state resonators, which are coupled to their material environments, the atoms here are isolated \textit{in vacuo}, and the thermal ``bath'' for the measured collective oscillation consists of the remaining collective modes of atomic motion.  After raising the trap potential to its final strength, the bath has an initial temperature of $2.9~\mu\mathrm K$, corresponding to a mode occupation of 0.54 phonons.

The atomic motion scatters cavity photons via collective virtual electronic excitations, causing transitions between collective motional energy levels (\textit{i.e.}, collective Raman scattering).  In this process, the collective atomic oscillator can absorb a mechanical energy quantum, thereby downconverting the emitted photon by $\wm$ (Stokes scattering).  If the oscillator is not in its ground state, it can alternatively upconvert the photon by $\wm$, thereby losing an energy quantum (anti-Stokes scattering).  By Fermi's golden rule, the rates for upconversion and downconversion, from an oscillator in its $\nu$th excited state, are proportional to $|\langle \nu | \hat \Z | \nu\mathord-1\rangle |^2 \propto \nu$ and $|\langle \nu | \hat \Z | \nu\mathord+1\rangle |^2 \propto \nu+1$, respectively.  A ground-state oscillator ($\nu=0$) will thus scatter only to the Stokes sideband, with more symmetric scattering indicative of higher phonon occupation.

The Fourier spectrum $n(\w)$ of photons emitted from the cavity, at a frequency $\w$ relative to the probe, and normalized by the discrete Fourier window time, can be calculated by solving a set of quantum Langevin equations \cite{marquardt:prl:07,safavi-naeini:11,supplemental}, yielding
\begin{multline}
\label{eq:psd}
n(\w) = \frac{\Com}{2}\frac{\k^2}{\k^2+\w^2} \\
\times \left [ \frac{\Gm^2\: \nubar}{(\w-\wm)^2+\Gm^2/4} + \frac{\Gm^2\:(\nubar+1)}{(\w+\wm)^2+\Gm^2/4} \right ],
\end{multline}
where $\nubar$ is the oscillator's mean phonon occupation.
Here we have parameterized the photon scattering rate by the dimensionless optomechanical cooperativity \cite{teufel:nature:11} $\Com \equiv 4\nbar \gom^2/\Gm \k$, which combines $\gom$, the mechanical damping rate $\Gm$, the mean intracavity probe photon number $\nbar$, and the cavity half-linewidth $\kappa$ ($2\pi\mathop\times 1.82~\mathrm{MHz}$).  In our atoms-based optomechanical system, $\Com \propto \nbar/\dca^2$, and the cooperativity can be tuned over many orders of magnitude by varying the probe intensity and $\dca$.

We measure the Stokes asymmetry by integrating the optical power $P_\pm$ scattered to frequencies near $\pm \wm$, detected using a balanced heterodyne receiver.  The detector measures a power spectral density $S_\mathrm{het}(\w) = S_\mathrm{SN}\left [ 1 + \varepsilon (n_0(\w) + n(\w))/2 \right ]$, where $S_\mathrm{SN}$ is the mean shot-noise spectrum as measured by the detector and $n_0(\w)$ is the spectrum of technical noise, which contributes less than $1\%$ to the observed spectrum at $\pm \wm$.  The quantum efficiency for measuring intracavity photons is $\varepsilon = 16\% \pm 2\%$. For each run of the experiment we measure the spectral density $S_A$ with atoms and $S_0$ without atoms (which measures $n_0$), and then measure $S_\mathrm{SN}$ by extinguishing the probe beam.  The unitless photon spectrum is then $n(\w) = 2(S_A-S_0)/\varepsilon S_\mathrm{SN}$.  In order to avoid excessive accumulated heating of the atomic gas, we acquire data for only 5~ms each run of the experiment, after which we observe the mechanical resonance to broaden significantly.

Measured spectra are shown in Fig.~\ref{fig:thermometry}.  At the lowest optomechanical cooperativites (\textit{viz.}, a weak probe detuned far from atomic resonance), the probe contributes only minimally to the phonon occupation.  Rather, the collective mode occupation should approach the average thermal occupation ($0.54\pm0.02$ phonons, measured via time-of-flight thermometry).  We measure a large Stokes asymmetry:  $P_-/P_+=3.0 \pm 0.8$, corresponding to a phonon occupation $\nubar = P_+/(P_--P_+)=0.49 \pm 0.10$.  We note that such a ``Stokes thermometer'' is self-calibrating, in that no experimental parameters are involved in extracting the phonon occupation number.

The mechanical resonance widths of the observed spectra are generally larger than the true mechanical damping rate, which we find by probing to the blue of cavity resonance and finding the phonon lasing threshold \cite{vahala:nphys:09,supplemental}, at which point the mechanical damping rate is equal to the optomechanical amplification rate.  In our system, we find a damping rate of $2\pi\times 1.5~\mathrm{kHz}$.  Comparing to the observed inhomogenous linewidth of $2\pi \times 3~\text{to}~4~\text{kHz}$ indicates that $\hat \Z$ dephases into approximately 2 to 3 collective normal modes, each containing 1,500 to 2,000 atoms.  We suspect that this dephasing is due mostly to slight differences in trap curvature in adjacent potential minima arising from the probe beam \cite{purdy:prl:10}, and partly to the anharmonicity of the standing-wave optical trap.

We next apply the sideband thermometer to probe the effects of measurement backaction.  A quantum measurement of position must be accompanied by a corresponding motional disturbance \cite{clerk:rmp:10}.  In  this experiment, where $\wm\ll\k$, we expect an increase in the phonon occupation by an amount $\Com/2$.  As we increase the cooperativity, the sideband asymmetry decreases as theoretically predicted.  Finally, we can compare the collective mode's mean phonon occup\-ation \mbox{$\nubar = \left \langle ( \sum_i g_i z_i/\gom ) ^2 \right \rangle / 2 \zho^2 - \sfrac 12$} to an upper bound on the mean single-atom phonon occupation $\nubar_a = \sum_i \left \langle z_i^2 \right \rangle / 2 N_a \zho^2 - \sfrac 12$, extracted by measuring the gas temperature using time-of-flight thermometry.  The upper bound is derived by assigning the gas's entire temperature increase after $1$~ms of probing (equivalent to many motional equilibration times) to the mean single-atom motion along the cavity axis.  The large discrepancy observed between the optically detected and single-atom occupations (Fig.\ \ref{fig:thermometry}b) highlights the fact that our detector senses the collective motion of the gas, rather than the motion of individual atoms.

The optical spectrum moreover serves as a record of the energy exchanged between light and motion.  A photon recorded at frequency $\w$ indicates the emission of an energy $\hbar\w$ from the atoms into the cavity field.  The spectral density of energy-exchange is $\hbar\w\, n(\w)$ (units of W/Hz), and the total power passing from the light to motion is \cite{marquardt:prl:07,murch:nphys:08}
\begin{equation}
	\label{eq:exchange}
	\nonumber
	P_\mathrm{om} = \frac{1}{2\pi} \int \hbar \w\, n(\w) d\w \approx \frac{\Gm \hbar\, \wm \Com}{2}\frac{\k^2}{\k^2+\wm^2}.
\end{equation}
Heat-exchange spectra for several values of the cooperativity, as well as heating rates taken from integrating over the sidebands, are shown in Fig.~\ref{fig:exchange}.  The heating rates agree well with the prediction of measurement backaction.  We additionally correlate the total energy exchanged with the atoms to the increase in the (finite) bath's temperature.  The bath temperature should increase by an amount given by equating the backaction heat with the temperature integral of the gas's heat capacity, which we calculate using Bose-Einstein statistics for axial motion and the ideal gas law for radial motion.  We find quantitative agreement to this theory as we vary both the cooperativity and the probe duration.

In this work, we have demonstrated the quantization of the collective motion of thousands of atoms, observing Stokes asymmetry and zero-point motion.  The Stokes asymmetry provides a self-calibrating thermometer for low-occupation collective modes.  We have in addition observed the spectrum of energy exchanged between light and collective atomic motion, spectroscopically identifying backaction heating from a quantum position measurement.   While our system measures center-of-mass motion, other modes \cite{safavi-naeini:11,brennecke:s:08,purdy:prl:10} of physical interest could be addressed by tailoring the light-motion interaction.  For example, quadrupole \cite{stringari:prl:96,altm07,cao:science:11} or scissors \cite{mara00} modes could be sensed using quadratic optomechanical coupling  \cite{purdy:prl:10}, allowing for precise measurements of the effects of interactions, superfluidity, and viscosity in degenerate Bose and Fermi gases.  Sideband spectroscopy of phonon modes in spatially extended gases \cite{brennecke:s:08} provides the means to study thermodynamics in static and driven systems \cite{baum10dicke}.

The authors acknowledge support from the AFOSR and the NSF.

\putbib[om]
\end{bibunit}

\begin{bibunit}[apsrev4-1]
\setcounter{equation}{0}
\setcounter{figure}{0}
\renewcommand{\theequation}{S\arabic{equation}}
\renewcommand{\thefigure}{S\arabic{figure}}

\section{Supplemental Material}

\paragraph{Photon spectrum:} For a single measured normal mode, an optomechanical system can be described by the Hamiltonian
\begin{equation}
\hat H = \hbar \wcav \dop{\hat a} \hat a + \hbar \wm \dop{\hat b} \hat b + \gom \dop{\hat a} \hat a (\dop{\hat b}+\hat b) + \hat H_\mathrm{in/out}.
\end{equation}
Here $\wcav$ is the cavity resonance frequency, $\hat a$ is the cavity field operator, and $\hat b$ is the mechanical field operator.

We solve the system's dynamics in the Fourier domain, writing $\tilde{O}(\w)$ to represent the Fourier transform -- with Fourier bandwidth $\wbw$ -- of an operator $\hat {O}(t)$.  We find the spectrum $n(\w) \equiv \wbw \langle  \tilde a_\mathrm{out}^\dagger(\w) \tilde a_\mathrm{out}(\w) \rangle$ of photons exiting the cavity by solving the system's quantum Langevin equations in the Heisenberg picture \cite{marquardt:prl:07,clerk:rmp:10,botter:arxiv:11}:
\begin{align}
\nonumber -i\w \tilde a(\w) &= \sqrt{2\k}\, \tilde a_\mathrm{in}(\w) - \k \tilde a(\w) \\ &\quad\quad - i \gom \sqrt{\nbar} (\tilde b^\dagger(-\w) + \tilde b(\w)), \\
\nonumber -i\w \tilde b(\w) &= \sqrt{\Gm}\, \tilde b_\mathrm{in}(\w) - i \wm \tilde b(\w) - \Gm \tilde b(\w)/2 \\ &\quad\quad - i \gom \sqrt{\nbar} ( \tilde a^\dagger(-\w) + \tilde a(\w) ),
\end{align}
where $\tilde a_\mathrm{in}(\w)$ and $\tilde a_\mathrm{out}(\w)$ are the cavity input and output fields, respectively, and $\tilde b_\mathrm{in}(\w)$ represents the mechanical bath.  Using the cavity output boundary condition $\hat a_\mathrm{in} + \hat a_\mathrm{out} = \sqrt{2\k}\, \hat a$, assuming $\Gm\mathop \ll \wm$, and applying Parseval's theorem ($\sum \dop{\tb} \tb \, \wbw \mathop= \sum \dop{\hat b} \hat b \, \Delta t$, for sampling interval $\Delta t$) yields Eqn.~(1) of the main article.

\begin{figure}[t]
	\centering
	\includegraphics[width=0.6\linewidth]{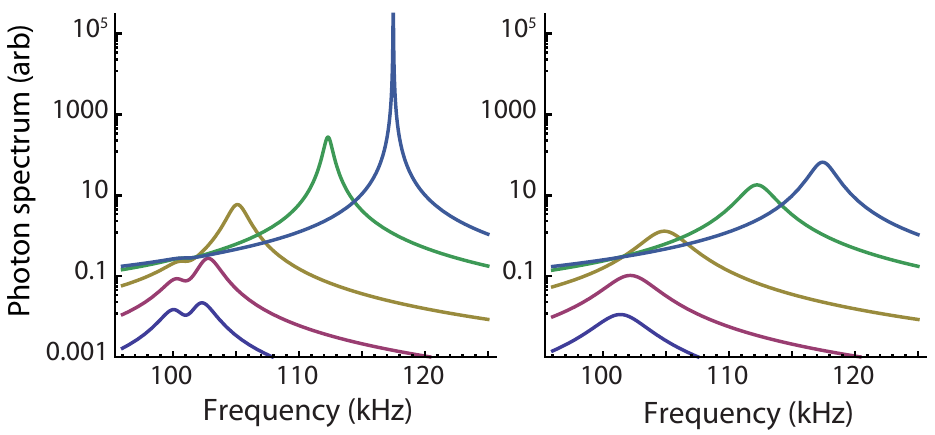}
	\caption{Analytic evaluation of the photon spectrum when probing at $\dpc=\k$.  Two cases, each having the same total optomechanical cooperativity, are considered:  A system with two normal modes (left), each damped at $\Gm=2\pi\times2~\text{kHz}$ and also dephasing at a rate $2\pi\times 2~\text{kHz}$, and a system with one normal mode (right), damped at $\Gm=2\pi\times 4~\text{kHz}$.  Each are plotted for increasing photon number $\nbar = 0.01,0.03,0.1,0.3,0.55$, with $\gom = 2\pi \sqrt{N_a} \times 3~\text{kHz}$.  For each, the spring shift gives the total optomechanical coupling, and the phonon lasing threshold gives the homogenous damping rate.}
	\label{fig:twoosc}
\end{figure}

\paragraph{Damping rate:} In order to measure the native damping rate of the collective atomic motion, we measure the system's phonon lasing threshold.  This is done by probing the cavity a detuning $\dpc = \k$ above the cavity's optical resonance.  Under this condition, the optomechanical spring \cite{marquardt:prl:07,botter:arxiv:11} shifts each normal mechanical mode's resonance frequency to a new value $\ws$.  In the limit $\wm\ll \k$,
\begin{equation}
\rule{0pt}{1.8em}
\ws = \sqrt{\wm^2 + 4\nbar \gom^2 \dpc / (\k^2+\dpc^2) }.
\end{equation}
The spring shift is also accompanied by an optomechanical amplification rate arising from cavity feedback, equal to 
\begin{equation}
\Gamma_\mathrm{amp} = \frac{2\k}{\k^2+\dpc^2} \left( \ws^2-\wm^2 \right ) = \frac{8\nbar \gom^2 \dpc \k}{ (\k^2+\dpc^2)^2}.
\end{equation}
When this amplification rate equals the native damping rate of the system, the system crosses a ``phonon lasing'' threshold; for $\Gamma_\mathrm{amp}>\Gm$, the oscillation becomes unstable and grows dramatically.  $\Gm$ can therefore be found by tuning $\ws$ until the system crosses the lasing threshold.  Fig.~\ref{fig:twoosc} shows, by solving the Langevin equations for one- and two-mode mechanical systems, how this technique can be used to differentiate between a system with a single measured normal mode and a system with multiple measured normal modes.

\putbib[om]
\end{bibunit}

\end{document}